\documentclass[conference]{IEEEtran}

\IEEEoverridecommandlockouts

\usepackage{cite}
\usepackage{amsmath,amssymb,amsfonts}
\usepackage{algorithmic}
\usepackage{graphicx}
\usepackage{textcomp}
\usepackage{xcolor}
\usepackage[ruled,vlined,linesnumbered]{algorithm2e}
\usepackage{booktabs}
\usepackage{array}
\usepackage{multirow}
\usepackage{enumitem}
\usepackage{makecell}  
\usepackage{bbm}
\usepackage{amsmath}

\def\BibTeX{{\rm B\kern-.05em{\sc i\kern-.025em b}\kern-.08em
    T\kern-.1667em\lower.7ex\hbox{E}\kern-.125emX}}
\begin{document}

\title{Blockage-Aware Non-stationary Dynamic Bandit for User Association in mmWave V2X Networks}


\author{\IEEEauthorblockN{
Weiqi Chi\IEEEauthorrefmark{1}, 
and
Manabu Tsukada\IEEEauthorrefmark{1}
}
\IEEEauthorblockA{\IEEEauthorrefmark{1}Graduate School of Information Science and Technology, The University of Tokyo, Japan}
Email: \{weiqichi, mtsukada\}@g.ecc.u-tokyo.ac.jp}

\maketitle 

\begin{abstract}
In millimeter-wave (mmWave) vehicular networks, dense base station (BS) deployments expand the user association (UA) decision space while dynamic blockages cause link quality fluctuations, posing critical challenges for effective mobility management. Traditional Multi-Armed Bandit (MAB) frameworks assume stationary reward distributions and fail to handle the rapid context-reward mapping shifts caused by vehicle mobility and transient blockages. To address this, we propose Blockage-Aware Non-stationary Dynamic Bandit (BAND), a fully distributed, channel state information (CSI)-free mobility management framework for mmWave vehicular networks, formulating UA as a non-stationary contextual bandit problem, enabling online adaptive optimization without requiring central coordination or offline training. BAND employs a cumulative sum-based change detection (CUSUM-CD) to dynamically narrow the active BS set, reducing exploration overhead while tracking reward distribution shifts. Proactive blockage detection suppresses transient signal degradation in the reward estimation process. Simulations demonstrate over 40\% regret reduction and up to 33.1\% network communication rate improvement compared with hypercube-based contextual bandit baselines, with robustness validated across varying blockage rates and network configurations.
\end{abstract}


\section{Introduction}
\label{intro}
Vehicle-to-Everything (V2X) communication is a cornerstone of intelligent transportation systems, demanding gigabit-level data rates and ultra-low latency. Millimeter-wave (mmWave) technology provides the spectrum abundance necessary for real-time sensor data exchange in vehicular networks. However, mmWave signals are highly susceptible to blockage by buildings, vehicles, and pedestrians, causing severe attenuation and sudden outages \cite{ruiz2021}. Vehicle mobility also introduces continuous topology changes, making user association (UA) a central component of mobility management in V2X mmWave networks, with direct impact on load balancing, spectrum efficiency, and quality of service \cite{uasurvey2016}. UA decisions must be made in real time to support safety-critical applications in vehicular networks, yet mobility-induced channel variation, intermittent blockage, and competing objectives make this a highly challenging problem in practice.

Traditional UA methods~\cite{docomo2010, RSSbased2017} depend on frequent channel state information (CSI) acquisition and centralized coordination, resulting in excessive signaling overhead and poor adaptability under high vehicle mobility~\cite{handoff}. In contrast, reinforcement learning (RL) approaches, particularly multi-armed bandit (MAB) frameworks, have gained attention for UA problems~\cite{RLV2Xsurvey}, which balance exploration-exploitation tradeoffs during in a online learning manner. In the MAB framework, vehicles act as agents selecting BSs (i.e., arms) and receive rewards reflecting communication quality at each time step. Due to fluctuating mmWave channels, reward distributions become non-stationary, motivating contextual bandits (CMAB) to learn context-reward mappings~\cite{onlinelearning}. Traditional CMAB partitions the context space into hypercubes, assuming vehicles within each hypercube receive similar rewards. However, this method suffers from high-dimensional context spaces where sparse samples within hypercubes significantly slow down the learning process. An alternative approach to tackle reward non-stationarity is to employ change-detection (CD) algorithms to monitor reward distribution shifts, including passive methods of Discounted Upper Confidence Bound (D-UCB)~\cite{dcucb2006}, Sliding-Window UCB (SW-UCB)~\cite{swucb2008}, active methods of Page-Hinkley Test (PHT)~\cite{hartland}, and cumulative sum (CUSUM)~\cite{liu2018change}. Nevertheless, these CD-based approaches face fundamental challenges in mmWave vehicular networks. Passive methods struggle to track rapid reward shifts across large-scale BSs under high mobility, while active methods are prone to false alarms by misidentifying reward degradation brought by transient dynamic blockages as persistent change points, triggering unnecessary algorithm resets and exploration.

In this paper, we propose Blockage-Aware Non-stationary Dynamic Bandit (BAND), a fully distributed CMAB-based approach that addresses non-stationary mmWave vehicular UA through dynamic BS set management and context-reward mapping shift tracking. The key contributions are fourfold: 
1) BAND enables practical mobility management in high-mobility vehicular networks by eliminating the need for prior CSI acquisition and offline training, significantly reducing signaling overhead;
2) We integrate CUSUM-CD into mobility management for the first time, enabling non-stationarity detection across a large BS action space in a fully distributed manner, validated under realistic dense mmWave network simulations;
3) Rather than applying CD directly to raw rewards, we proactively suppress transient distortions caused by dynamic vehicular blockage, reducing false alarms and ensuring the CD module responds exclusively to genuine reward distribution shifts;
4) We propose a dynamic BS management mechanism tailored for dense mmWave BS deployments, which monitors per-BS reward trends to prune under-performing BSs, directly addressing the scalability challenge in vehicular network scenarios and accelerating algorithm convergence.

The rest of this paper is organized as follows. Section~\ref{model} and ~\ref{BAND} introduce the system model and the BAND algorithm. Section~\ref{result} and~\ref{conclude} provide numerical results and conclusions.

\section{System Model} 
\label{model}
We investigate a vehicular communication network with densely deployed mmWave BSs. This section presents the mobility model, channel model, and problem formulation.

\subsection{Mobility and Channel Model}
We consider a finite horizon $T \in \mathbb{N}$ with discrete time steps $t = 1, \ldots, T$. Vehicles enter the network via a Poisson arrival process and update states periodically. The network comprises a BSs set $\mathcal{J} = \{1, \ldots, j, \ldots, |\mathcal{J}|\}$, where $|\mathcal{J}|$~is the total number of BSs. Both BSs and vehicles are equipped with antenna arrays for mmWave communication. At each time step $t$, let $\mathcal{V}^t = \{1, \ldots, k, \ldots, {|\mathcal{V}^t|}\}$ denote the set of vehicles in the network. As each vehicle is associated with one BS, the channel gain between BS $j$ and vehicle $k$ is:
\begin{equation}
y_{k,j}^t = \mathbf{h}_{k,j}^t\mathbf{w}_{k,j},
\label{eq.1}
\end{equation}
where $\mathbf{h}_{k,j}^t$ and $\mathbf{w}_{k,j}$ denote the channel matrix and beamforming vector of vehicle $k$~and BS $j$, respectively. When vehicle $i$ simultaneously communicates with BS $j$ at time $t$, it introduces interference to the ongoing transmission toward vehicle $k$. Specifically, the interference channel coefficient experienced at BS $j$ due to vehicle $i$ is $\tilde{y}_{i,j}^t$. Due to orthogonal frequency allocation across BSs, only intra-cell interference is considered. 
Therefore, the total interference at BS $j$ while serving vehicle $k$ is aggregated over all vehicles that simultaneously communicate with BS $j$:
\begin{equation}
I_{k,j}^t = \sum_{i \in (\mathcal{V}^t \setminus k)} P_{v}\left|\tilde{y}_{i,j}^{t}\right|^2\mathcal{I}_{i,j}^t  + N_o W,
\label{eq.3}
\end{equation}
where $W$ denotes the bandwidth of BS $j$, $P_{v}$ represents the vehicular transmit power, $N_o$ is the noise power spectral density, and $\mathcal{I}_{i,j}^t = 1$ if vehicle $i$ associates with BS $j$, otherwise $\mathcal{I}_{i,j}^t = 0$. We consider both static blockage from buildings and dynamic blockage from surrounding vehicles. When the line-of-sight (LOS) link is blocked, we assume complete blockage and neglect scattering and diffraction effects.

\subsection{Optimization Problem}
Note that intra-cell interference in \eqref{eq.3} depends on the joint association of all vehicles, and is implicitly fed back to each vehicle through the reward signal rather than through explicit coordination. Based on \eqref{eq.1} and \eqref{eq.3}, the instantaneous data rate between vehicle $k$ and BS $j$ is:
\begin{equation}
R_{k,j}^{t} = W \log_2 \left(1 + \frac{P_{v} |y_{k,j}^t|^2}{I_{k,j}^t}\right).
\label{eq.4}
\end{equation}
Let $\boldsymbol{\eta}^{t}$ denote the vector of associated BS indices for all vehicles in $\mathcal{V}^t$:
\begin{equation}
\boldsymbol{\eta}^{t} \triangleq [\eta_{1}^{t}, \ldots,\eta_{k}^{t}, \ldots, \eta_{{|\mathcal{V}^t|}}^{t}].
\label{eq.5}
\end{equation}
The UA problem maximizes aggregate network throughput by determining the optimal association vector:
\begin{equation}
\begin{aligned}
\max_{\boldsymbol{\eta}^t} \quad & r(\boldsymbol{\eta}^{t}) = \sum_{{k} \in \mathcal{V}^t} R_{k, \eta_{k}}^{t} \\
\text{s.t.} \quad & \sum_{j=1}^{|\mathcal{J}|} \mathcal{I}_{k,j}^{t} = 1, \quad \forall k \in \mathcal{V}^t.
\end{aligned}
\label{eq.6}
\end{equation}
The constraint in~\eqref{eq.6} ensures each vehicle connects to exactly one BS at time $t$, while each BS may serve multiple vehicles, therefore resulting in interference. This problem is NP-hard due to non-convex nonlinear constraints and integer variables~\cite{Mlika2018}. Although the optimal association can be identified with complete CSI via exhaustive search, it becomes computationally prohibitive in dynamic vehicular networks.

\section{Blockage-Aware Dynamic BS Management Bandit}
\label{BAND}
Dynamic blockages during vehicle movement are often brief, yet even short-duration blockages can cause nearly 15–20 dB signal loss in V2X communications~\cite{beffect}. In mmWave vehicular UA, applying CD algorithms without context awareness fails to capture context-reward mapping shifts brought by rapid channel variations, while blockage-induced signal degradation causes CD to misinterpret transient reward drops as genuine distribution shifts, triggering false-alarm breakpoints and unnecessary algorithm resets. On the other hand, naively incorporating blockage status as an additional context dimension in traditional CMAB frameworks introduces a sample sparsity problem that slows convergence, since blockages occur less frequently within each fine-grained area. To address these challenges, this section presents the BAND algorithm, covering the system architecture, environment assumptions, and algorithmic description.

\subsection{System Architecture}
\label{IIIA}
We formulate the UA problem within vehicular networks as a contextual bandit problem, where each vehicle $k$ acts as an agent, selects a BS $j$ from the available BS set, and receives a reward $R^t_{k,j}$ reflecting communication quality. We model the environment as piecewise-stationary~\cite{piecestationary}, where reward distributions of arms remain stable within intervals but shift abruptly at unknown breakpoints. This captures the characteristics of vehicular scenarios where channels stay relatively stable within a coherence time range before rapid changes occur due to mobility or dynamic blockages. A breakpoint occurs at time $t$ if $\exists j \in \mathcal{J}$ such that $\bar{R}_{k,j}^t \neq \bar{R}_{k,j}^{t+1}$, where $\bar{R}_{k,j}^t$ is the expected reward of BS $j$ for vehicle $k$ at time $t$. We impose two practical assumptions for analytical tractability:

\textit{\textbf{Assumption 1} (Piecewise Stationarity):} The minimum interval between any two consecutive breakpoints exceeds~$|\mathcal{J}|M$~for some integer $M$.

\textit{\textbf{Assumption 2} (Detectability):} There exists a known parameter $\zeta > 0$ such that $\forall j \in \mathcal{J}$ and $\forall t \leq T-1$, if a breakpoint occurs, then $|\bar{R}_{k,j}^t - \bar{R}_{k,j}^{t+1}| \geq 3\zeta$.

Assumption 1 guarantees adequate samples for reward estimation of each BS before their distribution changes occur, facilitating reliable detection. Assumption 2 excludes negligible estimated mean variations, ensuring that changes are detectable in practice.

\begin{figure}[htbp]
    \centering
    \includegraphics[width=0.45\textwidth, trim={0cm 0cm 0cm 0cm}, clip]{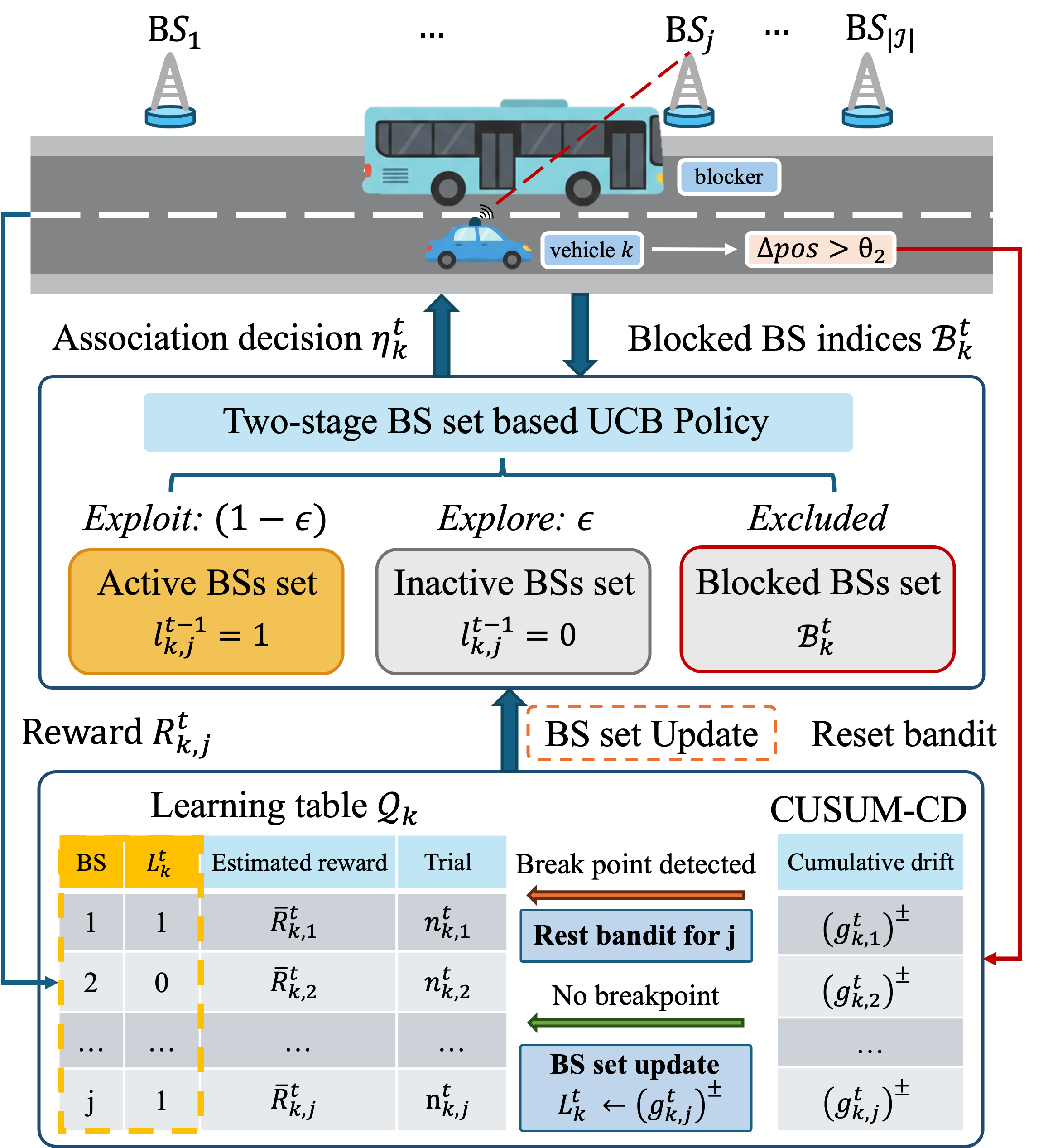}
    \caption{System framework}
    \label{fig:system}
\end{figure}

Fig.~\ref{fig:system} illustrates the BAND algorithm, which operates on each vehicle distributively through three steps. The system first identifies dynamic blockage status as context via geometry-based prediction. A two-stage upper bound confidence (UCB) policy is then applied. The first stage balances exploitation of active BSs with periodic exploration of inactive ones, and the second stage selects the BS with the highest UCB index for association. Finally, the system dynamically updates the BS set based on a two-sided CUSUM-CD algorithm, manage BS set according to their observed positive or negative cumulative reward drift. When a statistically significant reward drop is detected for BS $j$, the bandit algorithm is reset for that BS. When the vehicle detects a large position shift, a full algorithm reset of all BSs is triggered, implying an environmental context-reward mapping shift.

\subsection{Two-stage BS set-based UCB Policy}
\label{IIIE}
To balance exploration and exploitation while accounting for the time-varying nature of vehicular networks and dynamic blockage effects, we incorporate blockage status as context using the method from our 
previous work~\cite{vtcchi}, where dynamic blockages are detected with 
Fresnel zone radius $\tilde{r}=\sqrt{\lambda_c\frac{d_{bo} d_{vo}}{d_{bo}
+d_{vo}}}$. Here, $\lambda_c$ denotes the carrier wavelength, and $d_{bo}$ and $d_{vo}$ represent the distances from the BS and the receiver vehicle to the obstruction vehicle along the LOS path, respectively. Blockage occurs when an obstacle vehicle's height exceeds when 40\% of the Fresnel zone is obstructed. This geometric prediction is implemented as the \texttt{BlockageDetect} function described in Algorithm~\ref{alg:BAND}, which takes the position of vehicles and BSs as input and outputs the set of blocked BS indices $\mathcal{B}_k^t \subseteq \mathcal{J}$ for vehicle $k$ at time slot $t$. For tractability, any BS in $\mathcal{B}_k^t$ is assumed fully obstructed such that the link to vehicle $k$ is considered unavailable. With blockage status detected, vehicle $k$ carries out the following strategy for BS association:

\textbf{Stage 1: BS set Selection:} With probability $\epsilon$, the vehicle explores inactive BS set; otherwise, it exploits active BS set with probability $(1-\epsilon)$:
\begin{equation}
\mathcal{S}_k^t = \begin{cases}
\{j \in \mathcal{J} \setminus \mathcal{B}_k^t : l_{k,j}^{t-1} = 1\}, & \text{w.p. } (1-\epsilon), \\
\{j \in \mathcal{J} \setminus \mathcal{B}_k^t : l_{k,j}^{t-1} = 0\}, & \text{w.p. } \epsilon,
\end{cases}
\label{eq:setselection}
\end{equation}
where $\mathcal{S}_k^t$ is the selected BS set excluding blocked BS indices $\mathcal{B}_k^t$, and $\boldsymbol{L}_k^t = [l_{k,j}^t]_{j=1}^{|\mathcal{J}|}\in \{0,1\}$ denotes the BS status vector for vehicle $k$ at time slot $t$. Specifically, $l_{k,j}^t=1$ indicates that BS $j$ is categorized into the active BS set, otherwise it belongs to the inactive BS set. Blocked BSs are excluded as they are predicted to experience severe blockage with minimal probability of providing high-quality connections. The BS set is initialized by a BS-vehicle distance threshold $\theta_1$, reflecting the limited propagation range of mmWave communications.

\textbf{Stage 2: UCB policy:} Within the selected BS set $\mathcal{S}_k^t$, vehicles associate with the BS with the highest UCB index:
\begin{equation}
\eta_k^t = \arg\max_{j \in \mathcal{S}_k^t} \left\{ \bar{R}_{k,j}^t + c\sqrt{\frac{\ln t}{n_{k,j}^t}} \right\},
\label{eq:armselect}
\end{equation}
where $\bar{R}_{k,j}^t$ is the estimated reward for BS $j$ at time step $t$, $n_{k,j}^t$ is the number of trials on associating with BS $j$, and $c > 0$ is the exploration parameter of the UCB policy. This two-stage blockage-aware selection strategy prioritizes exploration among active BSs that are likely to provide high-quality connections, while periodically exploring inactive arms to detect potential changes in BS activity status. 

\subsection{BS set Management with Change Detection}
\label{IIID}

To adapt to non-stationary mmWave channel conditions in a fully distributed manner, BAND integrates a two-sided CUSUM-CD module that operates independently at each vehicle. This module monitors per-BS reward distribution shifts to trigger timely bandit resets upon environmental changes, while simultaneously adjusting the BS sets to condense the exploration space for faster convergence.

\subsubsection{\textbf{Two-sided CUSUM-CD}}
Under Assumptions 1 and 2, we adopt the two-sided CUSUM-CD algorithm for each BS $j$~\cite{liu2018change}, which monitors both positive and negative reward deviations from a baseline. We consider the first $M = 3$ reward samples collected by vehicle $k$ from BS $j$ before the breakpoint. The baseline is obtained by $\bar{\mu}_{k,j} \triangleq (\sum_{m=1}^{M} R_{k,j}^{m}/M)$. $(g_{k,j}^{t})^{\pm}$ records the two-sided cumulative drift of vehicle $k$ with respect to BS $j$ at time step $t$:
\begin{equation}
\begin{aligned}
(g_{k,j}^{t})^{\pm} = \max\left(0,\ (g_{k,j}^{t-1})^{\pm} \pm \left(R_{k,j}^t - \bar{\mu}_{k,j}\right) - \zeta\right),
\end{aligned}
\label{eq:CDalg}
\end{equation}
A breakpoint is detected when $(g_{k,j}^{t})^{\pm} \geq \tau$, where the detection threshold $\tau$ is for detection delay and false alarm rate balancing. Upon detection, the bandit is reset corresponding BS. Meanwhile, vehicles reset bandit algorithm for all BSs when there's a significant position shift exceeding threshold ($\Delta{\text{pos}_k} > \theta_2$).

\subsubsection{\textbf{BS Set Update}}
The BS set is dynamically updated according to the cumulative drift $(g_{k,j}^{t})^{\pm}$ from BS $j$. For vehicle $k$ at time slot $t$, we maintain a distributed learning table $\boldsymbol{\mathcal{Q}_k^{t}}$ defined as:
\begin{equation}
\boldsymbol{\mathcal{Q}}_k^{t} = \left\{ \boldsymbol{\bar{R}}_k^t, \boldsymbol{N}_k^t, \boldsymbol{L}_k^t \right\},
\end{equation}
where $\boldsymbol{\bar{R}}_k^t = [\bar{R}_{k,j}^t]_{j=1}^{|\mathcal{J}|}$ denotes the estimated reward vector, and $\boldsymbol{N}_k^t = [n_{k,j}^t]_{j=1}^{|\mathcal{J}|}$ denotes the trial count vector across BSs. The BS set management operates through two phases:
\begin{itemize}
    \item \textbf{Initialization:} When vehicle $k$ enters the network without historical knowledge, or when the whole bandit algorithm is reset, the BS set is initialized by filtering BSs based on the distance between the vehicle's current position and each BS. Only BSs within a predefined distance threshold are labeled as active. Then, $\boldsymbol{\bar{R}}_k^t$ and $\boldsymbol{N}_k^t$ are initialized as $\mathbf{0}$. 

    \item \textbf{CD-based BS set update:} During the UA process, the BS set status $l_{k,j}^t$ is updated based on the cumulative drift $(g_{k,j}^{t})^{\pm}$. An active BS $j$ is demoted to inactive if a negative cumulative drift is detected and its most recent reward falls below $\bar{\mu}_{\text{active}}$, the mean estimated reward across all currently active BSs. Conversely, an inactive BS $j$ is immediately promoted to active upon detection of a positive cumulative drift.
\end{itemize}

These designs collectively reflect the key properties of the BAND algorithm. Distance-based initialization exploits the limited propagation range of mmWave communications to confine early exploration to geometrically feasible BSs, accelerating the accumulation of observations required for reliable CD-based detection. The asymmetric promotion and demotion mechanism further enhances responsiveness, where cautious demotion prevents premature exclusion under transient fluctuations while immediate promotion encourages timely re-exploration of recovering BSs. By concentrating on the best-performing BSs, the UCB policy naturally feeds CUSUM-CD with the richest reward observations where context-reward mapping shifts manifest earliest, reducing adaptation latency as an emergent structural property of the joint algorithm without additional design overhead. Unlike hypercube-based CMAB approaches, the CD-triggered bandit reset mechanism offers flexible adaptability to context-reward mapping shifts and operates in a fully distributed manner on each vehicle without requiring centralized coordination and prior CSI acquisition. Furthermore, since BAND perceives the environment entirely through the CD algorithm, isolating blockage-affected BSs from the observation is essential. This operation ensures that CUSUM-CD responds to genuine reward distribution shifts rather than blockage-induced distortions. Finally, a formal algorithmic description of the BAND algorithm, detailing the integration of blockage detection, two-stage BS-level based UCB policy, and BS-level management, is provided in Algorithm~\ref{alg:BAND}.

\begin{algorithm}[t]
\caption{BAND}
\label{alg:BAND}
\SetAlgoLined
\SetInd{0.3em}{0.3em}
\textbf{Input:} $T$, $\mathcal{V}^{t}$, $\mathcal{J}$, $\boldsymbol{\mathcal{Q}}^0$\;
\For{$t = 1$ to $T$}{
    \For{each vehicle $k \in \mathcal{V}^{t}$}{
        \If{$k \notin \mathcal{V}^{t-1}$ \textbf{or} $\Delta pos_k > \theta_2$}{
            Initialize $\boldsymbol{\mathcal{Q}}_k^{init}$\;
        }
        $\boldsymbol{\mathcal{B}_k^t} \gets \texttt{BlockageDetect}$\;
        BS set $\boldsymbol{\mathcal{S}
        }_k^t$ selection according to~(\ref{eq:setselection})\;
        $R_{k,\eta_k^t}^{t} \gets$ Associate to BS $\eta_k^t$ according to~(\ref{eq:armselect})\;
        Calculate $(g_{k,j}^t)^{\pm}$ according to~(\ref{eq:CDalg})\;
        \scalebox{1}{$\boldsymbol{\mathcal{Q}}_k^t \begin{cases}
        \mathcal{Q}_{k,j}^{init}, & \text{if } (g_{k,j}^{t})^{\pm} > \tau \\
        l_{k,j}^t = 0, & \text{if } (g_{k,j}^{t})^{-} > 0,\ R_{k,j}^t < \bar{\mu}_{\text{active}} \\
        l_{k,j}^t = 1, & \text{if } (g_{k,j}^{t})^{+} > 0
        \end{cases}$}\;
    }
}
\textbf{Output:} associated BS indices set $\boldsymbol{\eta}$
\end{algorithm}

\subsection{Regret Analysis}
\label{IIIF}
Cumulative regret quantifies how much performance an algorithm sacrifices by not consistently selecting the best option $(\eta_{k}^{*})$. The system's expected regret over $T$ time steps can be expressed as~\cite{regret}:
\begin{equation}
\mathbb{E}[\mathcal{R}_{k}] = \max_{\eta_{{k}}^{t} \in \mathcal{J}} \mathbb{E}\left[\sum_{t=1}^{T} R_{k,\eta_{k}^{*}}^{t} - \sum_{t=1}^{T} R_{k,\eta_{k}^{t}}^{t}\right].
\end{equation}

For the computation of the expected regret in the following simulation, the optimal action is determined via an offline oracle that identifies the BS providing the highest instantaneous communication rate for vehicle $k$ at each time step $t$, based on the complete knowledge of channel realizations and interference levels. 

\section{Numerical Results}
\label{result}
We evaluated the performance of the proposed \texttt{BAND} algorithm in an urban mmWave vehicular network scenario spanning a simulated area of $550 \times 540$ $m$. To ensure realistic evaluation conditions, the simulation incorporates a dense deployment of mmWave BSs distributed throughout the region, combined with authentic vehicular traffic patterns generated using SUMO~\cite{SUMO2018}. The composition of simulated vehicle types adheres to the specifications in~\cite{3gpp_tr37885}, while the proportion of trucks is adjusted to emulate different dynamic blockage rates. The urban road topology and building infrastructure are derived from OpenStreetMap~\cite{osm2017} data for the Shibuya district in Tokyo, Japan. Channel characteristics are captured using the Clustered Delay Line (CDL) model integrated with ray tracing techniques~\cite{3gpp_tr38901}, enabling accurate representation of signal propagation dynamics influenced by static building obstructions. The complete set of simulation parameters is summarized in TABLE~\ref{tab:sim_params}. The algorithm parameters are selected through empirical testing across multiple parameter combinations.

\begin{table}[h]
\centering
\caption{Simulation Parameters and Settings}
\label{tab:sim_params}
\begin{tabular}{@{}lcc@{}}
\toprule
\textbf{Category} & \textbf{Parameter} & \textbf{Value} \\
\midrule
\multirow{7}{*}{\makecell{Network\\Setup}}
& Number of BSs & 69 \\
& Height of BSs & 5 m \\
& Simulation interval & 100 ms \\
& Carrier frequency & 28 GHz \\
& BS antenna size & $4 \times 4$ \\
& Vehicle antenna size & $2 \times 2$ \\
& Noise power spectral density & $-174$ dBm/Hz \\
\midrule
\multirow{4}{*}{\makecell{Traffic\\Parameters}}
& Simulation area & $550 \times 540$ m \\
& \multirow{3}{*}{\makecell{Vehicle\\Dimensions}} & $5 \times 2 \times 0.75$ m \\
& & $5 \times 2 \times 1.6$ m \\
& & $13 \times 2.6 \times 3$ m \\
\midrule
\multirow{4}{*}{\makecell{Algorithm\\Parameters}}
& C-UCB grid size & $10 \times 10$ m \\
& UCB exploration parameter $c$ & $\sqrt{0.5}$ \\
& BAND parameter ($\epsilon$, $\zeta$, $\tau$)  & (0.1, 0.05, 0.2) \\
& BAND distance threshold ($\theta_1$, $\theta_2$) & (200 m, 20 m) \\
\bottomrule
\end{tabular}
\end{table}

We compared our proposed algorithm with three baseline approaches and 2 ablation experiment: 1) a \texttt{C-UCB} algorithm that employs a CMAB-based UCB policy, partitioning the simulation region into hypercubes with stationary reward assumptions; 2) a non-learning \texttt{minDis} heuristic that always selects the nearest BS; 3) a near-optimal \texttt{maxRSRP} benchmark that associates with the BS of maximum RSRP given full CSI; 4) two ablation variants, \texttt{CUSUM-B} and \texttt{CUSUM-NB}, which adopt the BAND algorithm but disable dynamic BS set adjustment. Additionally, \texttt{CUSUM-NB} further disables blockage detection. Note that all algorithm parameters are empirically tuned based on the reward distribution and convergence behavior observed in simulation.

\begin{figure}[htbp]
    \centering
    \includegraphics[width=0.4\textwidth, trim={0cm 0cm 0cm 0cm}, clip]{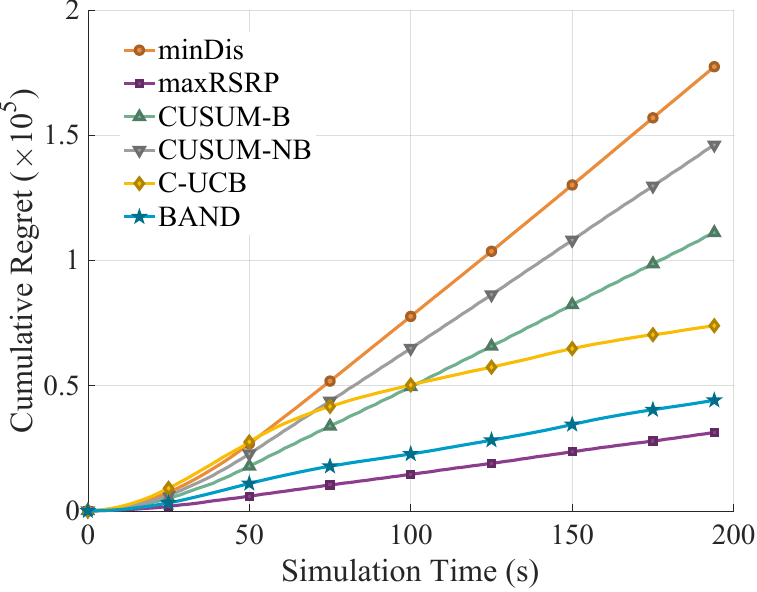}
    \caption{Cumulative regret comparison across learning time step}
    \label{fig:regret}
\end{figure}

Consider the vehicular network with a dynamic blockage rate of 30\%, where BSs operate on a 50 MHz bandwidth with 25 dBm transmission power. Fig.~\ref{fig:regret} compares the cumulative regret across learning time steps. Both online learning approaches (\texttt{BAND} and \texttt{C-UCB}) demonstrate convergence, while the non-learning \texttt{minDis} exhibits linearly increasing regret, confirming that distance-based heuristics fail to adapt to dynamic blockage-prone channels. The \texttt{maxRSRP} benchmark achieves near-optimal performance through exhaustive CSI collection from all BSs with the cost of signaling overhead. The proposed \texttt{BAND} algorithm significantly outperforms \texttt{C-UCB}, achieving 40.28\% regret reduction at 200 time steps. As shown in the figure, even with initial distance threshold $\theta_1$ for pruning, the absence of dynamic BS set management limits the ability of CUSUM-CD to adapt to environmental changes. Furthermore, excluding blocked BS to prevent CUSUM-CD from misidentifying breakpoints is shown to be necessary. BAND's performance benefits from dynamic and CD algorithms that rapidly adapt to evolving mmWave channel conditions with dynamic BS management. In contrast, \texttt{C-UCB}'s fixed hypercube partitioning leads to slow convergence due to sparse samples distributed across pre-defined regions.

\begin{figure}[htbp]
    \centering
    \includegraphics[width=0.5\textwidth, trim={0cm 0cm 0cm 0cm}, clip]{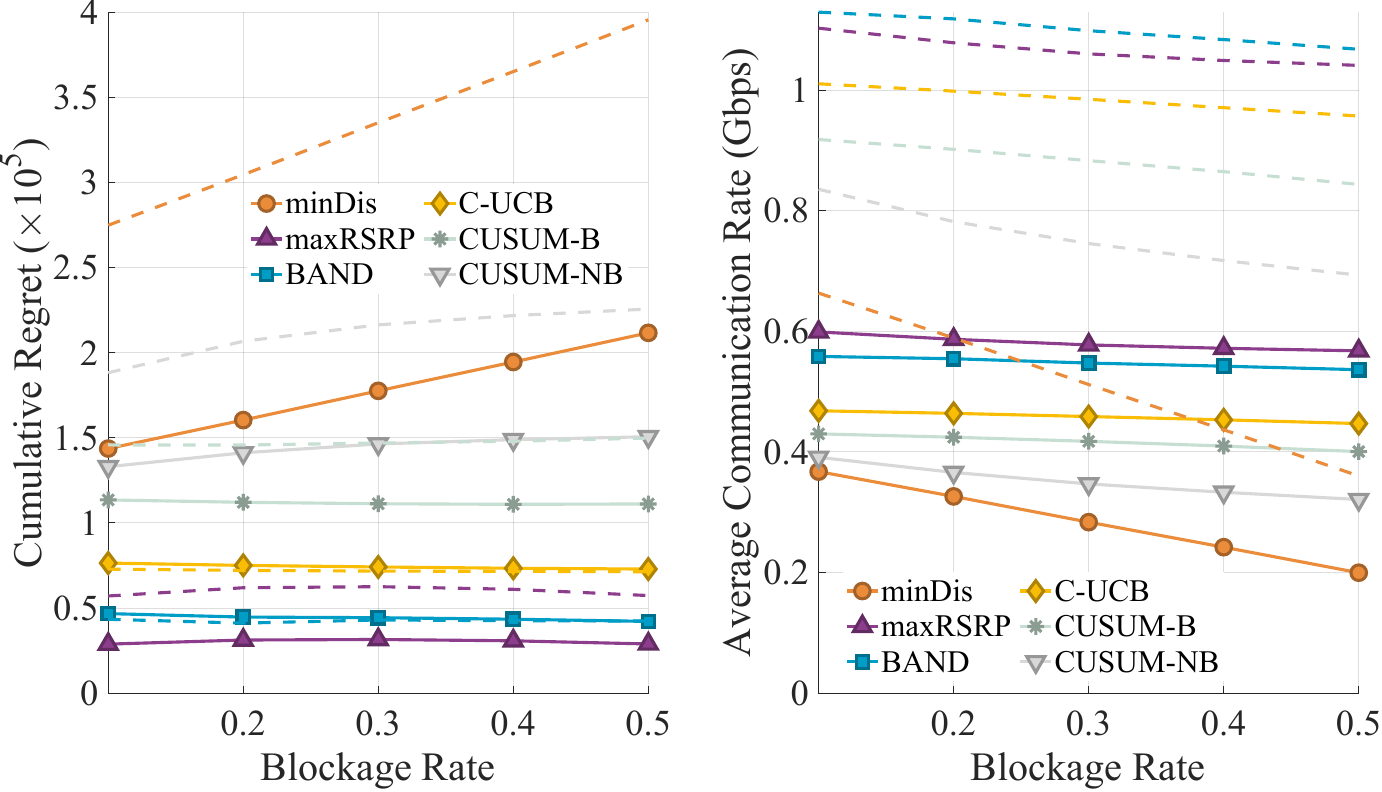}
\caption{Impact of blockage rate and bandwidth on algorithm performance: (a) Cumulative regret comparison, (b) Average communication rate comparison}
    \label{fig:crossbrate}
\end{figure}

Fig.~\ref{fig:crossbrate} illustrates the impact of blockage rate and bandwidth on algorithm performance. The solid and dashed lines of the same color represent the results of the same method under 50 MHz and 100 MHz bandwidth settings, respectively. \texttt{BAND} maintains consistently low cumulative regret across all blockage rates and bandwidth settings, with regret remaining nearly flat as blockage rate increases, while \texttt{minDis} and \texttt{CUSUM-NB} show substantial degradation. \texttt{BAND} also achieves stable average communication rates closely approaching the \texttt{maxRSRP} benchmark, confirming that the blockage detection mechanism effectively prevents transient blockage effects from contaminating reward estimation. \texttt{C-UCB} exhibits relative stability against blockage due to its fine-grained hypercube partitioning, which reduces the likelihood of blockage events within each partition. However, this comes at the cost of slower convergence, as evidenced by its higher overall regret. In contrast, BAND explicitly tracks context-reward mapping shifts via local position displacement $\Delta pos_k$, achieving effective adaptation at a coarser spatial granularity of 20m while attaining lower overall regret. Regarding bandwidth, while higher bandwidth improves communication rates, blockage-induced Line-of-Sight obstruction causes more severe rate degradation, further widening the gap between blockage-aware and non-aware algorithms.

\begin{figure}[htbp]
    \centering
    \includegraphics[width=0.5\textwidth, trim={0cm 0cm 0cm 0cm}, clip]{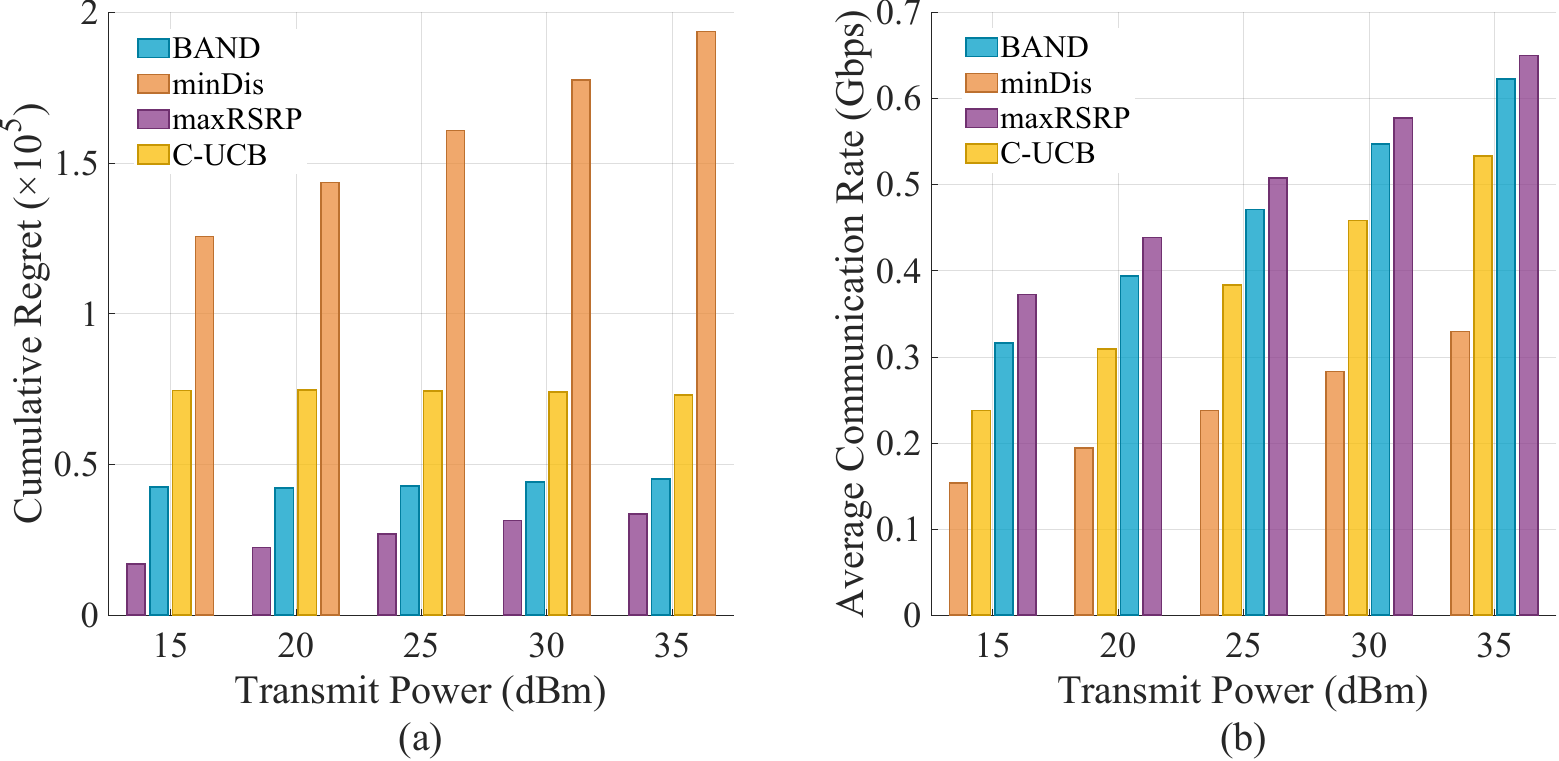}
\caption{Algorithm performance under varying transmission power (blockage rate = 30\%, bandwidth = 50 MHz): (a) Cumulative regret, (b) Average communication rate}
    \label{fig:bars}
\end{figure}


Fig.~\ref{fig:bars} demonstrates \texttt{BAND}'s consistent performance across varying vehicular transmit power, where ablation results are omitted for clarity. As shown in Fig.~\ref{fig:bars} (a), \texttt{BAND} consistently maintains the lowest cumulative regret, showing adaptability to fluctuating channel conditions compared to the \texttt{minDis} and \texttt{C-UCB}. Meanwhile, Fig.~\ref{fig:bars} (b) reveals that \texttt{BAND} achieves the higher communication rate, with a maximum 33.1\% improvement over \texttt{C-UCB} and a minimum 4.2\% gap to the \texttt{maxRSRP} upper bound. Furthermore, the performance gap narrows by nearly half as transmission power increases from lower to higher levels, indicating that \texttt{BAND} effectively exploits improved signal quality at higher transmission powers. These results confirm \texttt{BAND}'s robustness to different system configurations, demonstrating its ability to consistently identify high-quality BS associations while minimizing exploration overhead regardless of the underlying transmission power constraints.

Beyond performance metrics, the three approaches differ in the overhead required to reach each association decision. We exclude the association execution cost from this analysis, as it is identical across all approaches. At each time step, \texttt{maxRSRP} incurs $\mathcal{O}(T|\mathcal{J}|)$ signaling overhead through exhaustive CSI acquisition, while \texttt{C-UCB} introduces $\mathcal{O}(T|\mathcal{V}^t|)$ central communication rounds that scale with the number of vehicles. In contrast, BAND requires no CSI acquisition or central communication, as blockage detection and bandit reset rely solely on local position and BS location information, achieving zero additional communication overhead. 

\section{Conclusion}
\label{conclude}
This paper proposed a fully distributed BAND algorithm for user association in mmWave vehicular networks. By employing active change-detection to dynamically manage the candidate BS set and integrating proactively predicted blockage status as contextual information, the algorithm efficiently handled non-stationary rewards and transient blockages without requiring prior CSI. Simulation results demonstrated substantial performance improvements over benchmarks in both regret reduction and network transmission rate compared with the traditional CMAB framework-based algorithm, with robustness validated across different blockage rates. Future work will extend the framework to the joint optimization of user association to further improve spectral efficiency, alongside a more rigorous regret bound analysis.

\bibliographystyle{IEEEtran}
\bstctlcite{IEEEexample:BSTcontrol}
\bibliography{reference} 
\vspace{12pt}
\end{document}